\documentclass[reprint, superscriptaddress, amsmath,amssymb, aps,prd]{revtex4-2}

\RequirePackage[colorlinks=true
,urlcolor=magenta
,anchorcolor=blue
,citecolor=magenta
,filecolor=blue
,linkcolor=blue
,menucolor=blue
,linktocpage=true
,pdfproducer=medialab
,pdfa=true
]{hyperref}

\usepackage{amsmath,amssymb,amsthm,amsfonts,mathrsfs}
\usepackage{cprotect}
\usepackage{graphicx,tabularx}
\usepackage{rotating}
\usepackage{float}
\usepackage{newverbs}
\usepackage{multirow}  
\usepackage{cancel}
\usepackage{comment}
\usepackage{physics}
\usepackage{array}
\usepackage{enumitem}
\newcolumntype{P}[1]{>{\centering\arraybackslash}p{#1}}

\usepackage{xcolor}
\usepackage{dcolumn}
\usepackage{bm}

\newcommand{\DF}{\mathrm{DF}}
\newcommand{\Mach}{\mathscr{M}}

\begin{document}
\title{Echoes of Self-Interacting Dark Matter from Binary Black Hole Mergers} 

\author{Amitayus Banik}
\email{abanik@cbnu.ac.kr}
\affiliation{Department of Physics, Chungbuk National University, Cheongju, Chungbuk 28644, Korea}
\affiliation{Research Institute for Nanoscale Science and Technology, Chungbuk National University, Cheongju, Chungbuk 28644, Korea}

\author{Jeong Han Kim}
\email[Corresponding author: ]{jeonghan.kim@cbu.ac.kr}
\affiliation{Department of Physics, Chungbuk National University, Cheongju, Chungbuk 28644, Korea}

\author{Jun Seung Pi}
\email{junseung.pi@cbnu.ac.kr}
\affiliation{Department of Physics, Chungbuk National University, Cheongju, Chungbuk 28644, Korea}

\author{Yuhsin Tsai}
\email{ytsai3@nd.edu}
\affiliation{Department of Physics and Astronomy, University of Notre Dame, IN 46556, USA}

\begin{abstract}
Dark matter (DM) environments around black holes (BHs) can influence their mergers through dynamical friction, causing gravitational wave (GW) dephasing during the inspiral phase. While this effect is well studied for collisionless dark matter (CDM), it remains unexplored for self-interacting dark matter (SIDM) due to the typically low DM density in SIDM halo cores. 
In this work, by considering BH mergers within SIDM spikes, which can arise from models with a massive force mediator, we show that the GWs emitted are dephased in a distinct manner.
To incorporate the feedback of the BH orbital motion that can significantly modify the DM profiles, we use $N$-body simulations to analyze GW dephasing in binary BH inspirals within CDM and SIDM spikes. By tracking the binary's motion in different DM environments, we show that the Laser Interferometer Space Antenna (LISA) can observe GW dephasing arising from SIDM spikes in particular scenarios. Our results indicate that these observations offer a possibility of distinguishing between binary-BH inspirals in different DM environments.

\end{abstract}

\maketitle

\section{Introduction}
\label{sec:intro}
Gravitational wave (GW) observations~\cite{ LIGOScientific:2018mvr, LIGOScientific:2017vwq, LIGOScientific:2017ync, LIGOScientific:2016aoc, KAGRA:2023pio} have significantly enhanced our understanding of the universe, offering a powerful tool to study previously inaccessible physics~\cite{Baryakhtar:2022hbu, Adhikari:2022sve, Bertone:2019irm, AlvesBatista:2021eeu,Miller:2025yyx}. By analyzing GW signals with present and forthcoming detectors, such as the Laser Interferometer Space Antenna (LISA)~\cite{LISA:2022kgy, LISA:2017pwj, eLISA:2013xep, Barack:2003fp}, we can probe dark sector interactions beyond the Standard Model (SM) with remarkable precision. This approach enables the exploration of dark matter (DM) scenarios that evade direct detection via SM interactions, expanding the scope of DM search through purely gravitational effects.

Intermediate-mass black holes (BHs) with masses of $ \sim 10^2 - 10^5 M_{\odot} $ are believed to form in the centers of dwarf galaxies and globular clusters~\cite{Webb:2012cii}, with a single galaxy potentially containing multiple such BHs~\cite{Islam:2002gg, Bertone:2005xz, Rashkov:2013uua}. In particular, intermediate-mass BHs can host dense DM regions known as mini-spikes~\cite{Zhao:2005zr}. Such structures typically form via the adiabatic growth of small black hole seeds situated at the centers of DM halo~\cite{Quinlan:1994ed, Bertone:2005xz,Gondolo:1999ef}.
While the orbital motion of heavier BHs can disrupt their host DM spikes, this effect is milder for intermediate-mass BHs~\cite{Ullio:2001fb, PhysRevLett.88.191301, PhysRevLett.92.201304, Bertone:2005hw}. As a result, the surrounding DM structures may retain enough density to resist tidal disruptions, making it plausible that mini-spikes around intermediate-mass BHs have survived to the present day.

The persistence of these DM mini-spikes has significant implications for GW observations. Dynamical friction between DM mini-spikes and intermediate-mass BHs can perturb the orbits
in binary systems, significantly affecting their emitted GWs. The cumulative effects of
these interactions result in phase shifts in the gravitational waveforms, leading to 
observable deviations from the expected predictions without DM. This phenomenon, which 
we refer to as GW \emph{dephasing}, is sensitive to the profile of the DM spike and can thus
provide valuable insight into the DM distribution around intermediate-mass
BHs. Dephasing effects have been extensively studied in the context of collisionless dark matter (CDM)~\cite{Eda:2013gg, Macedo:2013qea, Barausse:2014tra, Eda:2014kra, Yue:2017iwc, Bertone:2019irm, Cardoso:2019rou, Hannuksela:2019vip, Kavanagh:2020cfn, Coogan:2021uqv, Cole:2022fir, Macedo:2013qea, Huang:2018pbu, Cole:2022yzw, Boudon:2023vzl, Aurrekoetxea:2024cqd,Cardoso:2019rou} and wave-like DM~\cite{Macedo:2013qea, Huang:2018pbu, Hook:2017psm, Cole:2022yzw, Kadota:2023wlm, Boudon:2023vzl, Aurrekoetxea:2024cqd, Croon:2017zcu,Choi:2018axi,Cao:2024wby,Blas:2024duy,Takahashi:2024fyq,Kim:2024rgf, Berezhiani:2023vlo, Banik:2025xwl}, including various astrophysical environmental effects~\cite{Barausse:2014tra,Barausse:2014pra}. In both cases, a sufficiently dense DM spike can significantly impact a black hole’s inspiral.

Self-interacting dark matter (SIDM) models~\cite{Arkani-Hamed:2008hhe,Feng:2009mn,Tulin:2013teo,Kaplinghat:2013yxa, Kaplinghat:2015aga}, where DM particles scatter off each other on timescales comparable to or shorter than halo formation, offer a compelling alternative to conventional collisionless dark matter (CDM). Many studies have shown how certain SIDM models can modify DM structures from dwarf galaxies ($\mathcal{O}(100)$~pc) to galaxy clusters ($\mathcal{O}(10)$~Mpc), addressing various small-scale structure challenges or seeding supermassive BHs (see~\cite{ Kamada:2016euw, Bullock:2017xww,Tulin:2017ara, Feng:2020kxv} for thorough reviews). The thermalization of SIDM through self-interaction leads to the formation of a lower-density SIDM spike~\cite{Kaplinghat:2015aga}, significantly altering the expected density when compared to a CDM spike. This difference directly impacts the GW dephasing, which, if detected, could provide crucial insights into SIDM spikes on much smaller scales ($\mathcal{O}(10^{-8})$~pc), comparable to $\mathcal{O}(10-100)$ of the Schwarzschild radius of the intermediate-mass BHs for generating visible LISA signals. 
While DM interactions have been explored in the context of facilitating supermassive black hole mergers~\cite{Alonso-Alvarez:2024gdz, NANOGrav:2024nmo}, no studies, to our knowledge, have investigated the dephasing signal in the SIDM scenario. This is likely due to the typically low spike density in SIDM, which results in negligible GW dephasing. In this article, we show that SIDM mediated by force mediators within a specific mass range can sustain a dense enough DM spike, leading to significant GW dephasing that is different from the CDM and no-DM scenarios. The SIDM scenarios arise when the higher DM velocity dispersion suppresses self-interactions within the spike, resulting in a denser profile than in typical SIDM scenarios.

Accurately modeling the interactions between the binary BH and the surrounding DM spike is essential for distinguishing different SIDM models through GW dephasing. In particular, during mergers 
involving intermediate-mass BHs, the energy injected into the DM spike can be 
comparable to its binding energy, leading to a feedback 
effect on the DM halo~\cite{Kavanagh:2020cfn, Mukherjee:2023lzn, Fischer:2024dte}, which can
alter the gravitational waveform. In this work, we use the $N$-body simulation code \verb|KETJU|~\cite{Rantala:2016rng, Mannerkoski:2019puf, Mannerkoski:2021hgr, Mannerkoski_2023} to study the evolution of binary-BH mergers, the surrounding DM spike, and the resulting GW signals. Defining the mass ratio between the lighter and heavier BH as $q$, we find for inspirals with $q = 10^{-4}$, our dephasing results match estimates based on Chandrasekhar’s formula for dynamical friction~\cite{Chandrasekhar:1943ys}. However, for $q = 10^{-2}$, we observe significant deviations due to changes in the DM halo, which reduce the dephasing. 

\section{DM Environments around BHs}
\label{sec:DMhalo}
The DM environment around a given binary-BH is determined by the constituent masses, commonly parameterized by the mass of the heavier BH $M_1$ and the mass ratio $q$. For simplicity, we fix $M_1 = 10^4\,M_{\odot}$ and consider cases with $q = 10^{-4}$ and $q = 10^{-2}$. The choice gives the Schwarzschild radius of the central BH as $r_s = 2 G M_1/c^2  \sim 10^{-9}$~pc ($c$ is the speed of light in vacuum, $G$ is Newton's gravitational constant). We highlight key features of the DM density profiles here, which we will use later to study the inspiral phase.

\subsection{Outer Region}

We assume the heavier BH resides at the galaxy's center. In the outermost regions, both CDM and SIDM halos follow the Navarro-Frenk-White (NFW) profile, $\rho_{\rm{NFW}}$~\cite{Navarro:1995iw, Maccio:2008pcd} given by:
\begin{align}
    &\rho_{\text{NFW}}(r) \equiv \frac{\rho_{\rm{sc}}}{\left(\frac{r}{r_{\rm{sc}}}\right)\left(1+\frac{r}{r_{\rm{sc}}}\right)^2}\,, 
    \label{eq:rho_NFW}\\
    &\quad\text{with} \quad \rho_{\rm{sc}} \equiv \frac{\rho_{\rm{cr}}\,\Delta}{3}\frac{c_{200}^3}{\ln(1+c_{200})-c_{200}/(1+c_{200})}\,. 
    \label{eq:rho_sc}
\end{align}
Here, $\rho_{\rm{cr}} = 2.775\times 10^{11}\,h^2\,M_{\odot}\rm{Mpc}^{-3}$ is the critical density for a spatially flat Universe with $h = 0.674$ being the value of the reduced Hubble constant today, as inferred from the CMB \cite{Planck:2018vyg}. We have defined the concentration parameter $c_{200} = r_{200}/r_{\rm{sc}}$ as the ratio between the virial radius and the scaling radius. Given the numerical parameter $\Delta = 200$, the mass enclosed within the virial radius, $M_{200}$, is related as:
\begin{equation}
    M_{200} = \frac{4\pi \rho_{\rm cr} \,\Delta}{3}r_{200}^3 \,.
    \label{eq:Mvir}
\end{equation}
Given a central BH mass $M_1$ (and the redshift $z$), one can obtain the virial (halo) mass~\cite{Girelli:2020goz, Behroozi:2019kql} and therefore $r_{200}$ through \eqref{eq:Mvir}. The technical procedure to obtain these quantities are described in Appendix~\ref{app:SIDM_density}. Furthermore, using relations between $c_{200}$ and $M_{200}$, such as those described in Refs.~\cite{Klypin:2014kpa, Correa:2015dva}, the scale radius $r_{\rm sc}$ and the scale density $\rho_{\rm sc}$ are obtained through the definition of the concentration parameter and \eqref{eq:rho_sc}, see Appendix~\ref{app:SIDM_density} for details. Therefore, these fix the parameters entering the NFW profile in \eqref{eq:rho_NFW}. To be precise, for $M_1 = 10^4\,M_{\odot}$, we obtain at $z = 0$, $M_{200} = 6.97 \times 10^9\,M_{\odot}$ corresponding to $c_{200} \approx 12$. This gives $r_{\rm sc} \approx 3779$~pc and $\rho_{\rm sc} \approx 6.66 \times 10^{-3}\,M_{\odot}\rm{pc}^{-3}$.

\subsection{CDM Spike}

\begin{figure}[t]
\centering
\includegraphics[width=0.48\textwidth]{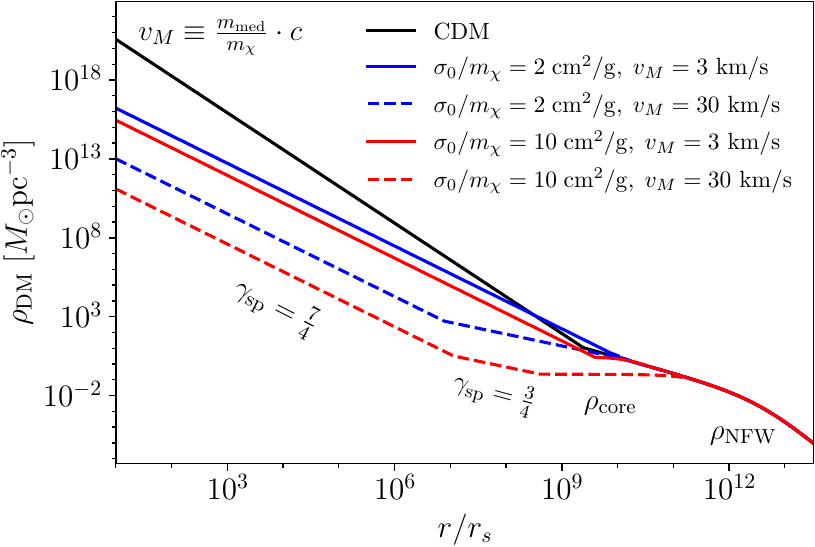}
\caption{Examples of SIDM density profiles with different $\sigma_0/m$ (colored lines) and $v_M$ (solid and dashed lines) around a black hole binary with $M_1 = 10^4\,M_{\odot}$ and $q=10^{-4}$. We have here $r_s \sim 10^{-9}$ pc. The solid black curve shows the corresponding CDM profile for comparison, while the dashed red curve highlights different regions of the SIDM profile. For numerical study, we use the CDM (solid black), solid blue, and solid red curves, with the binary inspiraling from $r\approx 30\,r_s$ and merging at $r \approx 10\,r_s$.
}
\label{fig:density_profile}
\end{figure}

For CDM, the BH's influence modifies the NFW profile, forming a dense spike~\cite{Gondolo:1999ef, Ullio:2001fb}, 
\begin{equation}
    \rho_{\rm{spike}}(r) \equiv \rho_{\rm{sp}} \left(\frac{r_{\rm{sp}}}{r}\right)^{\gamma_{\rm{sp}}}\,, \quad {\rm for} \quad r \leq r_{\rm sp}\,,
    \label{eq:rho_spike}
\end{equation}
where we take the spike index $\gamma_{\text{sp}} = 7/3$, assuming an adiabatically grown CDM spike~\cite{Gondolo:1999ef}. The parameters in the spike profile to be determined are the spike radius $r_{\rm{sp}}$ and spike density $\rho_{\rm{sp}}$. The spike radius is determined by $r_{\rm sp} \approx 0.2\, r_{\rm 2M}$~\cite{Merritt:2003qk}, where $r_{\rm 2M}$ is the radius up to which the mass enclosed by the NFW profile is twice the mass of the central BH, i.e, we have
\begin{align}
     4\pi&\int_{r_{\rm{min}}}^{5 r_{\rm{sp}}} dr \,r^2\,\rho_{\rm{CDM}}(r) = 2M_1\,,
     \label{eq:CDM_cond}
\end{align}
where for $r > r_{\rm sp}$, $\rho_{\rm CDM} = \rho_{\rm NFW}$ and for  $r \leq  r_{\rm sp}$, we have $\rho_{\rm CDM} = \rho_{\rm spike}$. Here, $r_{\rm{min}} = 10\,r_s$, which as described in the next section, is the separation distance at which the binary is defined to merge. Finally, continuity between the NFW profile and the spike profile,
\begin{equation}
    \rho_{\rm spike}(r_{\rm sp}) = \rho_{\rm NFW}(r_{\rm sp})\,,
\end{equation}
fixes $\rho_{\rm sp}$. Therefore, transitioning from the NFW profile to the spike profile, we have the corresponding spike parameters given by $r_{\rm sp} = 2.25$~pc and $\rho_{\rm sp} = 11.16\,M_{\odot}{\rm pc}^{-3}$.

\subsection{SIDM Core and Spike}

For SIDM, we estimate the density profile following~\cite{Alonso-Alvarez:2024gdz}. Self-interactions thermalize SIDM particles allowing them to achieve hydrostatic equilibrium, forming a core of radius $r_c$, with the NFW profile for $r \geq r_c$~\cite{Kaplinghat:2015aga}. The formation of the core depends on the strength of the interaction cross-section satisfying \cite{Kaplinghat:2015aga, Alonso-Alvarez:2024gdz}
\begin{equation}
    \frac{\langle \sigma v \rangle}{m_\chi}\,\rho_{\text{NFW}}(r_c)\,t_{\text{age}} \sim 1\,,
    \label{eqn:cond}
\end{equation}
i.e., approximately one scattering per DM particle within the core, $r < r_c$, during the age of the core chosen to be $t_{\text{age}} = 0.5$~Gyr \footnote{The effect of different core ages has been studied in the context of supermassive BHs in Ref.~\cite{Alonso-Alvarez:2024gdz}, considering $t_{\rm age}\sim0.1-1$~Gyr. Increasing $t_{\rm age}$ leads to a wider core (with fixed $\sigma_0/m_\chi$ and $v_{M}$); for e.g. increasing $t_{\rm age}$ from 0.5 to 1 Gyr for $(\sigma_0/m_\chi,\,v_M) = (10\,\rm cm^2/g,\,30\,km/s)$ increases $r_c$ from 320 pc to 740 pc, which can be compensated by increasing $\sigma_0/m_\chi$ to $20\,\rm{cm}^2/g$. All our results can therefore be scaled for larger or smaller $t_{\rm{age}}$.}. Here, $\langle \sigma v \rangle$ is SIDM's averaged velocity times cross-section, and $m_\chi$ is the DM mass. We model the interaction between DM through the exchange of a particle of mass $m_{\text{med}}$. Introducing a threshold velocity, $v_{M} \equiv  c\cdot (m_{\text{med}}/m_\chi)$, the velocity-dependent cross-section behaves as:
\begin{equation}
    \sigma(v) = \frac{\sigma_0}{1+(v/v_{M})^4} 
    =
    \begin{cases}
       \sigma_0\,, \quad v \ll v_{M} \equiv \frac{m_{\text{med}}}{m_\chi}\,,  \\[1mm]
        \sigma_0 \left(\frac{v_M}{v}\right)^4 \,,\quad v \gg v_{M} \,.
    \end{cases}
    \label{eq:sigma_vel}
\end{equation}
Therefore, for DM velocities smaller than $v_M$, the interaction is effectively contact-type, and for much larger velocities, it behaves as if through the exchange of a massless mediator. To obtain the SIDM core profile $\rho_{\rm core}$, we solve the Poisson equation $v_0^2 \nabla^2 \ln \rho_{\rm core}(r) = -4\pi G\rho_{\rm core}(r)$, where $v_0$ is the average dispersion velocity in the core. Due to the equilibrium state achieved by the DM, $v_0$ is constant within the isothermal core. Since self-interactions simply redistribute particles inside the core, we require the SIDM core mass to match the NFW mass within $r \leq r_c$~\cite{Alonso-Alvarez:2024gdz}. Along with continuity at $r = r_c$ with the NFW profile, these conditions define $\rho_{\rm core}$. The full details on obtaining the SIDM core are described in Appendix.~\ref{app:SIDM_density}.

The SIDM spike follows a similar power-law behavior as \eqref{eq:rho_spike} as shown in Ref.~\cite{Shapiro:2014oha}. However, the exponent varies depending on the interaction type, with $\gamma_{\text{sp}} = 3/4 \,(7/4)$ for a contact-type (massless-mediator-type) self-interaction. In contrast to CDM, where the spike radius is determined by comparing the gravitational potential of the BH to that of the enclosed DM, cf. \eqref{eq:CDM_cond}, the spike radius of the SIDM can be estimated by matching the kinetic energy of particles in the isothermal profile right outside of the spike, to the virial velocity set by the central BH. With this, we follow Ref.~\cite{Shapiro:2014oha} to identify $r \leq r_0 = G M_1/ v_0^2$ as the radius beginning at which the DM motion is influenced primarily by the BHs and transitioning into a spike profile. Furthermore, Ref.~\cite{Shapiro:2014oha} also shows that, as a result of solving the governing hydrodynamic equations, there is a corresponding power law for the velocity, $v(r) \propto (r_0/r)^{1/2}$ for $r \ll r_0$, leading to a possible change in the interaction type, c.f. \eqref{eq:sigma_vel}, at a transition radius $r_t$ (specific details in Appendix.~\ref{app:SIDM_density}).

In Fig.~\ref{fig:density_profile}, we show the CDM profile (black) and 
examples of SIDM profiles. For a fixed $v_M$, increasing
$\sigma_0/m_\chi$ widens the SIDM core, as observed in the dashed-blue versus dashed-red curves. The wider core leads to suppressed densities in both the spike and core regions. From the core boundary to the halo center, the dashed curves transition from a flat core (contact interaction) to a contact-interaction-spike $(\gamma_{\rm sp} = 3/4$), and finally to a massless-mediator-spike ($r_t$, $\gamma_{\rm sp} = 7/4$). For heavy mediators (corresponding to $v_M = 30$~km/s), the SIDM spike densities are much lower than in the CDM case. However, with lighter mediators (solid blue and red, $v_M = 3$~km/s), $\sigma(v)$ experiences stronger velocity suppression, allowing a denser $\gamma_{\rm sp} = 7/4$ spike to form near the core boundary, leading to significantly larger spikes. 

Below, we focus on the $v_M = 3$~km/s scenario, where a $\sim 10$~pc core and a dense DM spike lead to significant GW dephasing. The formation of such DM spikes around intermediate-mass BHs of mass $10^{3-6}~M_{\odot}$ has been discussed in the literature~\cite{Zhao:2005zr}. Moreover, observations of dwarf galaxies with radius $\sim 10$~pc and $\mathcal{O}(1)$~km/s virial velocities~\cite{Pace:2018tin} suggest that halos with a similar size can exist and may host intermediate-mass BHs. In larger halos with velocity dispersions $\gg 3$ km/s, the self-interactions of the DM become velocity-suppressed, making the DM behave more like CDM. 

As a closing remark, we show in the next section that the dephasing of GWs due to the presence of DM halos is sensitive mainly to the DM density within the inspiral region, rather than the full profile. In scenarios where the shape of the DM profile varies, for example, from different modeling of the halo~\cite{Cole:2022fir}, we expect to observe similar signals if the intrinsic properties lead to similar central densities.

\begin{figure*}[t]
\centering
\includegraphics[width=0.78\textwidth]{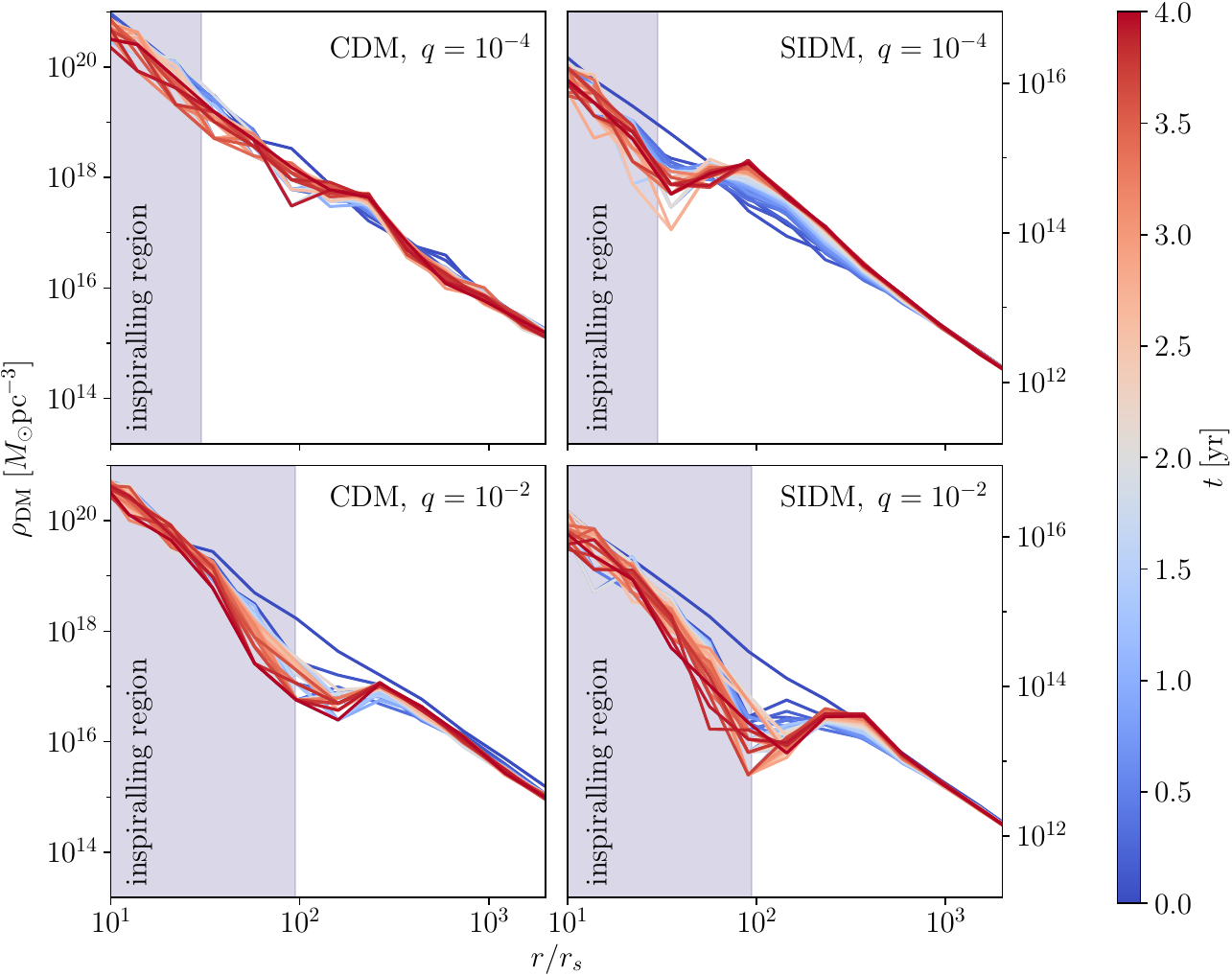}
\caption{The evolution of the SIDM density profiles around a BH of mass $10^4\,M_{\odot}$ over the binary merger time of $\sim 4$ years, for binary mass ratios $q = 10^{-4}$ (top row) and $10^{-2}$ (lower row). Purple-shaded regions refer to the inspiralling region for the respective binary. 
}
\label{fig:density_time}
\end{figure*}

\section{Binary-BH inspiral and GW dephasing}
\label{sec:inspiral}
In this study, we utilize the publicly available \verb|KETJU| code~\cite{Rantala:2016rng,Mannerkoski:2019puf,Mannerkoski:2021hgr,Mannerkoski_2023} to model the trajectories of binary-BHs and the resulting GW signals. \verb|KETJU| allows us to track the evolution of binary-BHs from the early stages of their merging process up to the final few orbits before their coalescence. This approach accounts for the dynamical friction exerted by the surrounding DM. Additionally, \verb|KETJU| incorporates velocity-dependent post-Newtonian (PN) corrections for binary-BHs, providing a more precise representation of their dynamics. In the simulations presented here, PN corrections are applied at 2.5PN order for the binary motion, ignoring the effect of precession.

The density profiles are used as initial conditions for the simulation. In Appendix.~\ref{app:code}, we provide details of our simulations and their implementation. We focus on the self-interaction strength $\sigma_0/m_\chi$ = 2 -- 10$\text{ cm}^2/\text{g}$ with $v_M = 3$ and 4 km/s.  Although our modified code includes the effect of self-interactions, we observe them to be inefficient during the observation time, $\tau_{\rm obs} = 4$~years, as can be estimated using 
\begin{equation}
    N_{\rm scatt} = \frac{\sigma_0}{m_\chi}\left(\frac{v_M}{v}\right)^4\,v\,\rho_{\rm ins}\,\tau\,,
    \label{eq:rate}
\end{equation}
where $\rho_{\rm ins}$ is the density at the start of the inspiral, see Fig.~\ref{fig:density_time}. Note that \eqref{eq:rate} is a generalization of the condition \eqref{eqn:cond} and gives the scattering rate per DM particle within the spike. Accordingly, we have taken the velocity-averaged cross-section \eqref{eq:sigma_vel} within the spike where DM velocities are up to $\mathcal{O}(0.1c)$. Given that $\rho_{\rm ins} \sim 10^{16}\,M_{\odot}{\rm pc}^{-3}$ for the cases of interest in Fig.~\ref{fig:density_profile}, this leads to $N_{\rm scatt}\sim 10^{-4}$ for $\tau = \tau_{\rm obs}$. Therefore, $\sigma_0/m_\chi$ essentially is used to determine the SIDM spike density, by setting the extent of the isothermal core, and we do not include self-interactions between DM particles in the \verb|KETJU| simulation.
Furthermore, as we have noted, heavier mediators result in lower densities; therefore, we restrict ourselves to this narrow range of $v_M$ to obtain observable dephasing effects. As demonstrated later, this places a bound on the mediator masses for SIDM arising from detectability by LISA. To account for numerical fluctuations, we performed 5 simulations per case using the same setup and similar initial conditions, presenting the averaged simulation results along with uncertainties estimated via $1/\sqrt{N}$ analysis. For uncertainties on the derived quantities, we accordingly propagate the uncertainties.

The binary merger proceeds through energy losses in the system. In the absence of a DM environment (i.e., in a vacuum), the merger occurs purely through GW emission~\cite{Peters:1964zz, Maggiore:2007ulw}. Having fixed $M_1 = 10^4\,M_{\odot}$ with a $q =\,10^{-4}\,\text{or}\,10^{-2}$, and given that, in our simulation, the binary is defined to merge at $r \approx 10\,r_s$, the frequency of the GW at merger is $\sim10^{-1}$~Hz, so that the inspiral frequencies cross the high sensitivity regime of LISA. We then proceed to set up the inspiral for a merger time of $\tau_{\rm{merge}} = 4$~years, corresponding to starting our simulation with the distance between the BHs being $\approx 30\,r_s\,(95\,r_s)$ for $q = 10^{-4}\,(10^{-2})$. This gives frequencies of the emitted GWs in the range of $10^{-3}$ to $10^{-1}$~Hz. However, if DM is present around the BHs, it gains energy from their orbital motion, imparting dynamical friction to the binary system and thus accelerating the merger rate~\cite{Chandrasekhar:1943ys}. Additionally, the binary's motion can disrupt DM density profiles, introducing more complex dynamics.

Fig.~\ref{fig:density_time} shows the evolution of the CDM and SIDM density profiles around binaries (in different colors) for two mass ratios. For $q = 10^{-4}$, the DM profiles remain largely stable, indicating insufficient orbital energy to disrupt them. At $q = 10^{-2}$, DM density near the initial inspiral radius (right edge of the purple region) drops by up to an order of magnitude, as the lighter BHs clear out DM along their trajectory. The disruption occurs near the region at the beginning of the inspiral, due to the fact that the secondary BH spends most of the duration of the merger in this region. In particular, we find that a duration of $\sim 0.5$~years is required to clear out the density at a given radius, as shown in Fig.~\ref{fig:density_fixedR} of Appendix~\ref{app:code}. In comparison, in the initial stages of the inspiral, approximately 2~years are required to reduce the orbital radius from $\sim 90\,r_s$ to $\sim 80\,r_s$, whereas in the final stages of the inspiral, the distance between the BHs reduces rapidly within 0.2~years (see Fig.~\ref{fig:a_vs_t} of Appendix~\ref{app:code}).

To study the effect of GW dephasing from dynamical friction due to DM,  we calculate the total number of orbit cycles accumulated over the merger time $\tau_f$,
\begin{equation}
    N^{\rm{acc}}_{\rm{cyc}} \equiv \int_{0}^{\tau_{f}} f(\tau) \,d\tau,\quad \tau_f \equiv\tau\big|_{r=10\,r_s}\,,
    \label{eq:Ncyc_acc}
\end{equation}
where $f$ is the instantaneous frequency of the GW signal. \verb|KETJU| solves the orbital equations of motion, providing information about the binary's position and velocity at each time step. This allows calculation of the orbital frequency~\cite{Peters:1964zz} at that particular time step, and correspondingly the GW frequency, which we use to obtain the accumulated orbital cycles in \eqref{eq:Ncyc_acc}.

For the vacuum case, we find $N^{\rm{acc}}_{\rm{cyc, vac}} = 2.67 \times 10^{6}$ and $ N^{\rm{acc}}_{\rm{cyc, vac}} = 4.96 \times 10^5$ for $q = 10^{-4}$ and $q = 10^{-2}$ respectively. As dynamical friction reduces the merger time $\tau_f<\tau_{\rm merge}$, fewer cycles are accumulated w.r.t. the pure vacuum case, and this is characterized by the \textit{dephasing}
\begin{equation}
    \Delta N_{\rm{cyc}} = N^{\rm{acc}}_{\rm{cyc},\,\rm{vac}} - N^{\rm{acc}}_{\rm{cyc},\,\rm{env}}\,\,.
    \label{eq:dephase}
\end{equation}
Here, the subscript ``env" refers to the orbital cycles accumulated for a binary inspiralling within a DM environment. In order to compare our dephasing results, based on $N$-body simulations, to static dynamical friction results, we make use of the publicly available code \verb|IMRIpy|~\cite{Becker:2021ivq, Becker:2022wlo}, which studies the time-evolution of the binary without incorporating feedback on the DM halo (see the Appendix~\ref{app:code} for details). For CDM, our simulation gives  $\Delta N_{\rm{cyc}} = 47302 \pm 21154$ with a BH mass ratio of $q = 10^{-4}$. The static dynamical friction estimate yields  $\Delta N_{\rm{cyc}} = 21787$, which is within $1.64\sigma$ of our simulation result, also being consistent with recent $N$-body simulations involving CDM, for e.g.~\cite{Mukherjee:2023lzn}. For a mass ratio of $q = 10^{-2}$, our simulation finds $\Delta N_{\rm{cyc}} = 576 \pm 258$, roughly two orders of magnitude smaller than the dynamical friction estimate $\Delta 
 N_{\rm{cyc}} = 58636$.

\begin{figure}[t]
\centering
\includegraphics[width=0.48\textwidth]{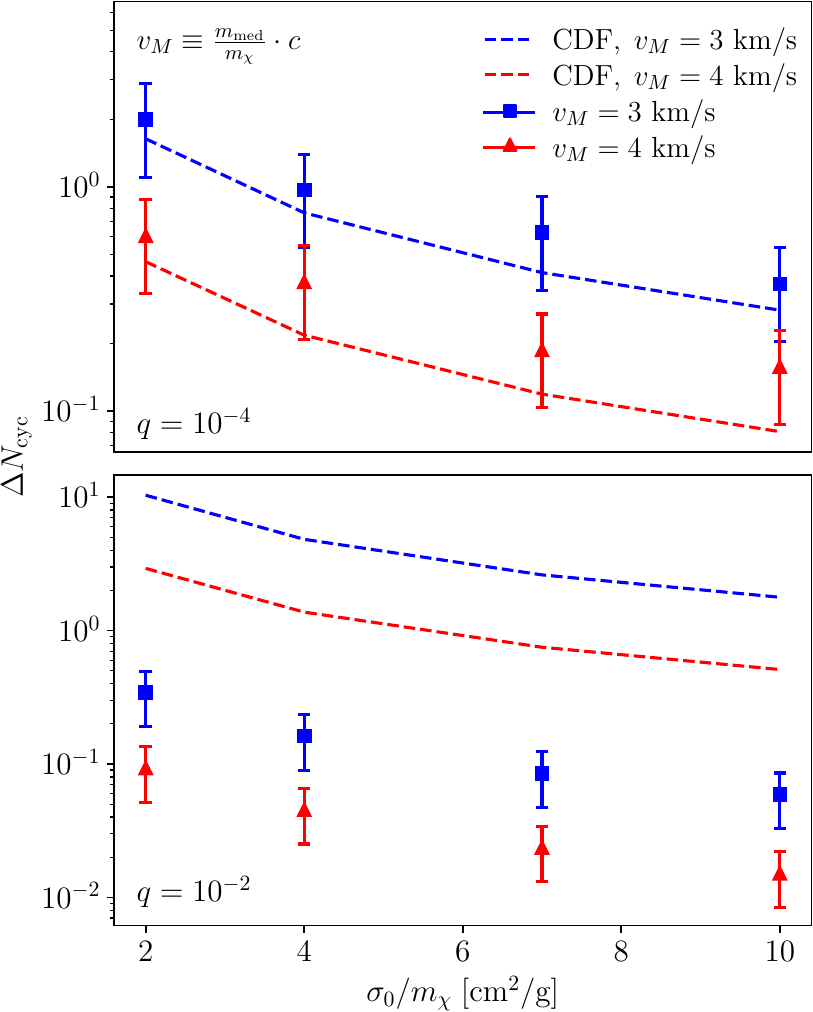}
\caption{The dephasing obtained for the SIDM environment from our simulation, for various $\sigma_0/m_\chi$, $v_M = 3$ km/s (blue squares) and 4 km/s (red triangles), and BH mass ratios $q = 10^{-4}$ (upper panel) and $10^{-2}$ (lower panel). We show the dephasing obtained using Chandrasekhar's formula (CDF), which assumes no feedback on the DM halo from the secondary BH (blue and red dashed lines). The error bars on the bullets are based on $1/\sqrt{N}$ analysis of the 5 simulations per case we performed.}
\label{fig:Ncyc_prob}
\end{figure}

Fig.~\ref{fig:Ncyc_prob} shows dephasing in SIDM environments for two BH mass ratios, two $v_M$ values, and varying $\sigma_0/m_\chi$. The data points (blue and red bullets) are from our simulation, and the dashed curves are static dynamical friction model estimates. As in the CDM case, $\Delta N_{\rm{cyc}}$ from the simulation is much smaller than the dynamical friction model estimate for $q = 10^{-2}$. This reduction in the dephasing for larger $q$ is expected, as the DM density in the inspiral orbit is cleared out more rapidly. Indeed, for $q = 10^{-2}$, the DM halo is significantly disrupted after about 0.37 years (see Fig.~\ref{fig:density_fixedR} in Appendix~\ref{app:code}),  causing the accumulation of $\Delta N_{\rm{cyc}}$ to effectively stall up to this point. We note that the dephasing, even for SIDM, is much larger than the $\Delta N_{\rm cyc}$ from possible baryonic effects estimated in Refs.~\cite{Barausse:2014tra,Barausse:2014pra}.

We comment briefly on the choice of the initial orbital radius and its impact on the calculated dephasing. As noted in Ref.~\cite{Kavanagh:2020cfn}, because we have initialized the second BH directly into a static DM distribution, a numerical transient can arise as the secondary initially experiences the unperturbed static density before depleting the local DM environment. In a realistic scenario, the secondary would begin the inspiral from a larger radius, carving out the DM profile before reaching the radius from which observation begins. To check the impact of this choice, we performed a test simulation of the $q = 10^{-2}$ binary system, where the feedback is prominent, now initialized at a larger radius. We still evaluate the dephasing for the time period $\tau_{\rm obs} = 4$~years from merger, as in our main simulations. We find that the DM halo again experiences a depletion close to the region of the initial radius, as in Fig.~\ref{fig:density_time}. By the time the secondary reaches the radius from which we begin tracking the dephasing, the local dynamical friction is now lower due to the reduced density. Overall, we find that this leads to a slight decrease in the dephasing. While in principle the true astrophysical initialization occurs much further out, thereby introducing an inherent uncertainty regarding the exact initial state of the halo profile, our test indicates that the effect of the choice of initial condition appears to be modest, as the total $\Delta N_{\rm cyc}$ is largely determined by the high-density inner region near the end of the inspiral.

Finally, we note that beyond dynamical friction and the gravitational forces of DM that we have incorporated through our $N$-body simulations, other effects such as the accretion~\cite{1939PCPS...35..405H,Bondi:1944rnk,Macedo:2013qea} of DM onto the secondary BH can affect the inspiral motion, and therefore the emission of GWs. Accretion increases the mass of the secondary BH over its orbits and results in an additional loss of orbital energy, thereby serving as another source of dephasing by reducing the time to merger. For the cases analyzed in this work, the corresponding merger times  $\tau_{\rm merge} \approx 4$~years may result in the effect of accretion on $\Delta N_{\rm cyc}$ being typically subdominant to dynamical friction, as noted in studies treating ultra-light DM~\cite{Kadota:2023wlm} and CDM spikes~\cite{Karydas:2024fcn}. However, for longer inspirals~\cite{Karydas:2024fcn}, the effects can become comparable, resulting in an overall larger dephasing.  

\section{Distinguishing DM Environments through GW Dephasing}
\label{sec:distinguish}
To access LISA's potential in distinguishing GW signals between CDM and different SIDM environments, we use the $N_{\rm{cyc}}$ template derived assuming no DM environment to fit the early- and late-time evolution of the $N_{\rm{cyc}}$ spectrum. By comparing the different template parameters obtained from the fit to the expected LISA sensitivity on these parameters, we can decide if the change of the inspiral evolution due to the DM environment is observable or not. 
 
As mentioned earlier, the $N$-body simulation results allows us to calculate $N^{\rm acc}_{\rm{cyc}}(f)$ as a function of emitted GW frequency, with lower (higher) frequencies corresponding to the inspiral at earlier (later) times. On the other hand, the vacuum template is given by ~\cite{Peters:1964zz, Maggiore:2007ulw, Moore:2016qxz}:
\begin{align}
    &N^{\rm acc}_{\rm{cyc, vac}}(f) = \frac{1}{32\pi^{\frac{8}{3}}}\,\left(\frac{G M_{\rm{ch}}}{c^3}\right)^{-\frac{5}{3}}\,\left[f^{-\frac{5}{3}}-f_{\rm{max}}^{-\frac{5}{3}}\right]\,,
    \label{eq:Ncyc_fit}
\end{align}
where $f_{\rm max}$ equals the GW frequency at merger time when $r=10\,r_s$.
The chirp mass is defined as $M_{\rm{ch}}=M_1 q^{3/5}/(1+q)^{1/5}$. For $ M_1 = 10^4\,M_{\odot} $ and $ q = 10^{-4} \,(10^{-2}) $, we get $ M_{\rm{ch}} \approx 39.81 \,M_{\odot} \,(629.70 \,M_{\odot})$. Therefore, as indicated by \eqref{eq:Ncyc_fit}, fewer orbital cycles are accumulated for a larger $M_{\rm ch}$, as confirmed by our aforementioned results of $N^{\rm{acc}}_{\rm{cyc, vac}}$. 

Fitting the simulation results for different DM environments using (\ref{eq:Ncyc_fit}) requires corrections $\delta M_{\rm{ch}}$ to match the data. Therefore, we allow $M_{\rm{ch}}$ to perturb by $\delta M_{\rm{ch}}$, which will be determined by a least-squares fit to $N^{\rm acc}_{\rm{cyc},\,\rm{env}}$, calculated using our simulation data. This relates the orbital cycles accumulated in the DM environment to those in vacuum as \eqref{eq:Ncyc_fit}:
\begin{align}
    &N^{\rm acc}_{\rm{cyc, env}}(f) = \frac{1}{32\pi^{\frac{8}{3}}}\,\left(\frac{G (M_{\rm ch}+\delta M_{\rm{ch}})}{c^3}\right)^{-\frac{5}{3}}\,\left[f^{-\frac{5}{3}}-f_{\rm{max}}^{-\frac{5}{3}}\right] \nonumber \\
    &\qquad \qquad \approx N^{\rm acc}_{\rm cyc,vac}(f) - \frac{5}{3}\left(\frac{\delta M_{\rm ch}}{M_{\rm ch}}\right) N^{\rm acc}_{\rm cyc, vac}(f)\,,
\end{align}
where we have retained the term first order in the relative shift in the chirp mass $(\delta M_{\rm ch}/M_{\rm ch})$. We can then estimate the \textit{relative} dephasing as:
\begin{equation}
    \frac{\Delta N_{\rm cyc}}{N^{\rm acc}_{\rm cyc, vac}} \equiv \frac{N^{\rm acc }_{\rm{cyc,vac}} - N^{\rm acc}_{\rm{cyc, env}}}{N^{\rm acc}_{\rm cyc, vac}} = \frac{5\,}{3}\left(\frac{\delta M_{\rm ch}}{M_{\rm ch}}\right)\,.
    \label{eq:dephasing_analytic}
\end{equation}
which gives an analytic insight through the dimensionless \textit{relative shift} $\delta M_{\rm ch}/M_{\rm ch}$. As indicated by \eqref{eq:dephasing_analytic}, if the dephasing $\Delta N_{\rm cyc}$ is fixed, $\delta M_{\rm ch}$ increases overall for a larger $q$, on account of the $N_{\rm cyc,vac}^{\rm vac}$ decreasing and $M_{\rm ch}$ increasing.

\begin{figure}[t!]
\centering
\includegraphics[width=0.48\textwidth]{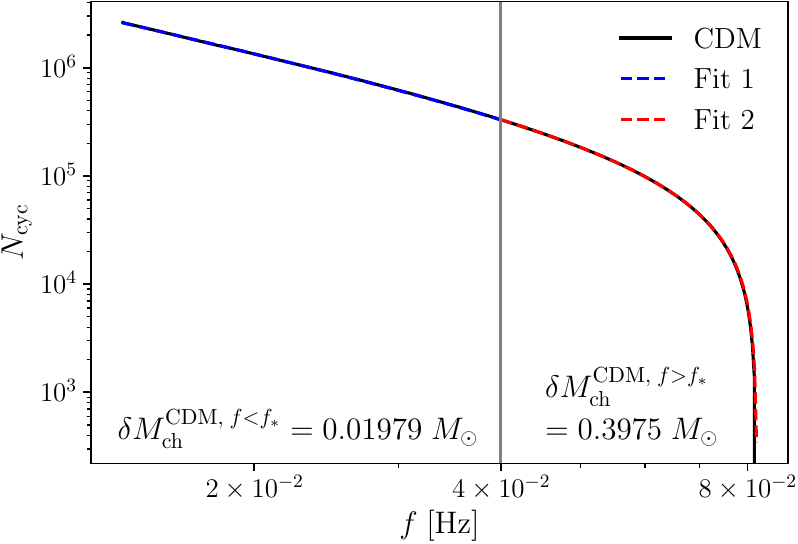}
\caption{An example of the fitting procedure employed for CDM data with $q = 10^{-4}$, from one of the five simulations we performed. The gray line separates the two fitting regions, where we calculate the indicated shifts before and after $f_* = 0.04$~Hz.}
\label{fig:fit_eg}
\end{figure}

To determine the shift in the chirp mass, we fit the spectrum in frequency regions smaller or larger than $f_*$, which is chosen to minimize the total chi-square of the fit, and consider the chirp mass difference:
\begin{equation}
    \delta M^{i}_{\rm{ch}} \equiv \delta M^{i,f<f_*}_{\rm{ch}} - \delta M^{i,f>f_*}_{\rm ch}\,,
    \label{eq:chirp_mass_diff}
\end{equation}
for $i=$ CDM, or SIDM models between the two frequency regions. Here, for the lower frequency region $f< f_*$, $f_{\rm max}=f_*$, whereas for $f >f_*$, $f_*$ equals the GW frequency at merger time when $r = 10\, r_s$. In this manner, without requiring phase measurements from an extremely early time to determine the difference in total $N_{\rm{cyc}}$, $\delta M^{i}_{\rm{ch}}$ allows us to distinguish between the early- and late-time evolution of the inspiral across different DM environments, which is not provided by the estimate \eqref{eq:dephasing_analytic}. An example of this is displayed in Fig. \ref{fig:fit_eg}, whereby we fit the vacuum template to the CDM environment.  The exact values obtained for these shifts are given in Tab. \ref{tab:uncertainties} of the App.~\ref{app:Fitting}. However, comparing the results of the relative shift from our fitting procedure with the analytic result in \eqref{eq:dephasing_analytic}, we find agreement within an order of magnitude. To estimate the errors on the shift in the chirp mass, we make use of \eqref{eq:dephasing_analytic}, by applying the method of error propagation, to obtain 
\begin{equation}
    \frac{\sigma_{\delta M_{\rm ch}}}{M_{\rm ch}} = \frac{3}{5}\left(\frac{\sigma_{\Delta N_{\rm cyc}}}{N_{\rm cyc, vac}^{\rm acc}}\right)\,.
     \label{eq:dephasing_err}
\end{equation}
Here, $\sigma_x$ denotes the error on the quantity $x$. For the error on the dephasing $\sigma_{\Delta N_{\rm cyc}}$, we use a $1/\sqrt{N}$ analysis as mentioned before, as $N_{\rm cyc}$ is derived from our simulation results. The other quantities $M_{\rm ch}$ and $N_{\rm cyc, vac}^{\rm acc}$ are fixed and therefore do not propagate in the error on the chirp mass shift. 

\begin{figure}[t]
\centering
\includegraphics[width=0.48\textwidth]{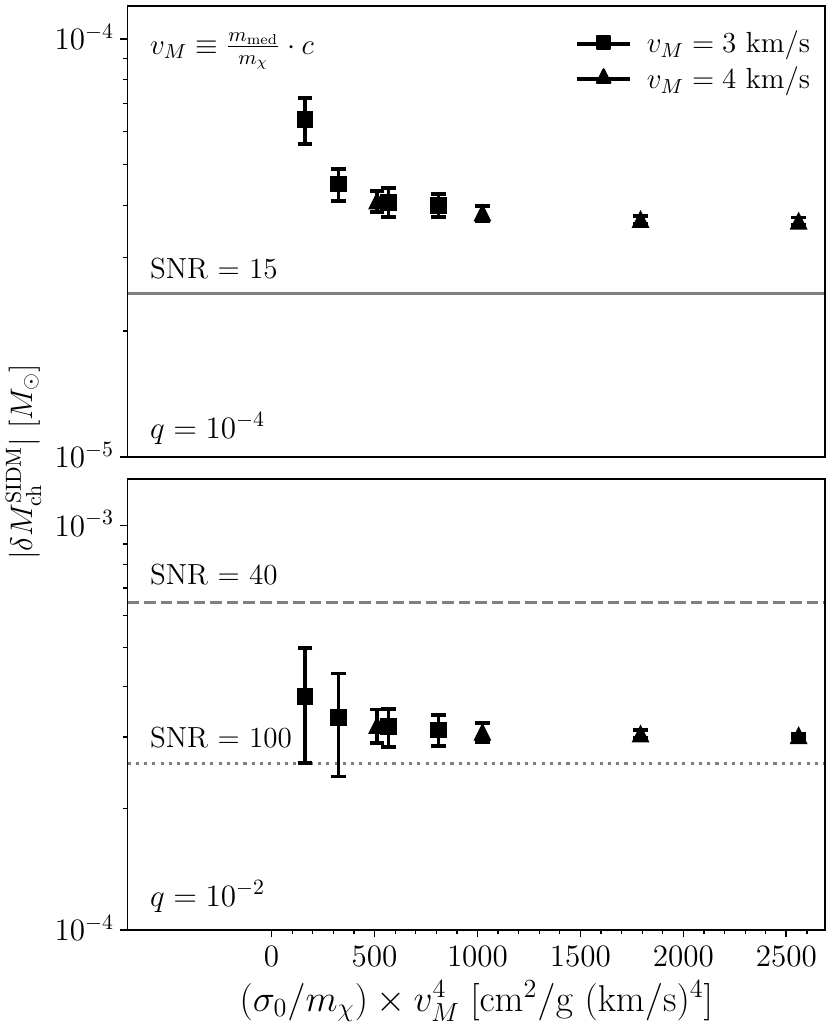}
\cprotect\caption{The shifts in the chirp mass for $N_{\rm cyc}$ measurements for $q = 10^{-4}$ \textit{(top)} and $q = 10^{-2}$ \textit{(bottom)} with as a function of the combination of SIDM parameters, $\sigma_0/m_{\chi}$ and $v_M$. Gray lines (solid, dashed and dotted) mark the uncertainties on the measurement of the chirp mass by LISA for various SNRs, as calculated using \verb|GWFish|. Squares (triangles) show the averaged results from the 5 simulations we performed, obtained from the sets with $v_M = 3\,(4)$ km/s, with the error bars obtained using the method of error propagation from \eqref{eq:dephasing_err}. For both mass ratios, the shifts are distinguishable from the vacuum template, but for the larger mass ratio, a higher SNR is required, on account of the feedback on the SIDM spike. The shifts decrease for increasing $\sigma_0\, v^4_M/m_{\chi}$ as this leads to spikes with lower densities. 
}
\label{fig:SIDM_delta}
\end{figure}

To access LISA's ability in distinguishing $N_{\rm{cyc}}(f)$ across DM environments, we check if $\delta M_{\rm ch}^{\rm CDM}$ and $\delta M_{\rm ch}^{\rm SIDM}-\delta M_{\rm ch}^{\rm CDM}$ can be larger than the expected LISA sensitivity of the $M_{\rm{ch}}$ measurement. Using \verb|GWFish| \cite{Dupletsa:2022scg}, a Fisher matrix-based tool to forecast parameter uncertainties in GW signals, we estimate the measurement uncertainty of $M_{\rm{ch}}$ by LISA. Given the merger time $\tau_{\rm merge} = 4$~years, we vary the luminosity distance from 20 -- 120 Mpc (500 -- 3500 Mpc) for $q = 10^{-4}\,(10^{-2})$, to forecast observations with SNRs 15, 40, and 100, obtaining the corresponding uncertainties on the chirp masses.

For the phase difference between the CDM and vacuum environments, we find  $|\delta M_{\rm ch}^{\rm CDM}| = (0.51 \pm 0.19)\,M_{\odot}$ for $ q = 10^{-4} $ and $ (0.14 \pm 0.063)\,M_{\odot} $ for $ q = 10^{-2} $. These differences surpass LISA’s sensitivity, even with a modest SNR of 15, capable of detecting  $\delta M_{\rm ch} \approx 2.5 \times 10^{-5} M_\odot $. Corresponding to SIDM, we find an average $|\delta M_{\rm ch}^{\rm SIDM}| \approx 4.2 \times 10^{-5}\,M_{\odot}$ for $ q = 10^{-4} $ and $|\delta M_{\rm ch}^{\rm SIDM}| \approx 3.1 \times 10^{-4}\,M_{\odot}$ for $q = 10^{-2}$ as shown in Fig.~\ref{fig:SIDM_delta} for the parameters $\sigma_0/m_\chi$ and $v_M$ characterizing the SIDM profile. As studied in Sec.~\ref{sec:inspiral}, the dephasing depends only on the density of the spike profile, which is determined by these SIDM parameters, as discussed in Sec.~\ref{sec:DMhalo}. Therefore, the mass shift can be characterized as a function of a combination of the SIDM parameters $(\sigma_0/m_\chi)\times v_M^4$. This reveals a degeneracy for the SIDM parameters, for example, the nearly overlapping markers for $(\sigma_0/m, v_M) = (7\,{\rm cm^2/g}, 3\,{\rm km/s})$ and $(\sigma_0/m, v_M) = (2\,{\rm cm^2/g}, 4\,{\rm km/s})$ leading to a $\delta M_{\rm ch} \approx 4 \times 10^{-5}\,M_{\odot}$, arising from SIDM spikes with similar densities. Finally, for $ q = 10^{-2} $, the SIDM spike is significantly disrupted, causing dephasing to end earlier, requiring a higher SNR$\sim 100$ in order for LISA to resolve the small $ \Delta N_{\rm cyc} $ differences from vacuum. As we show, the SIDM spikes considered can generate an observable dephasing signal that can be distinguished from inspirals in CDM and vacuum. Finally, we note that despite the limited statistics of our $N$-body simulations, the chirp-mass shifts of the different SIDM scenarios lie above the LISA sensitivity. Improving the simulation statistics would allow one to investigate whether LISA could distinguish SIDM scenarios with different model parameters, which we leave for a future study.

\section{Conclusion}
\label{sec:conclude}

DM surrounding binary BHs affects merger dynamics by inducing dynamical friction, leading to GW dephasing during the inspiral phase. This dephasing depends on the DM spike profile, which can be altered by self-interactions. While SIDM typically forms a core that reduces spike density, we show that velocity-dependent SIDM with a massive mediator can sustain a dense enough spike, for certain regions of the parameter space, to produce a detectable dephasing signal at LISA. Using $N$-body simulations, we analyze GWs from binary-BH inspirals in both CDM and SIDM, tracking dephasing through chirp mass shifts derived from fitting early- and late-time inspiral motion to a vacuum template. Our results indicate that the SIDM scenario can indeed generate an observable dephasing of the GW signal, different from the CDM and no-DM cases, for a particular combination of the mediator mass and interaction strength. This suggests that intermediate-mass BH inspirals can offer a new way to probe SIDM structure on much smaller scales than those typically considered in dwarf galaxies.

\section*{Acknowledgments}
We would like to thank Peter H. Johansson for his valuable comments and for providing the publicly available \verb|KETJU| code. 
AB, JHK and JP are supported partly by the National Research Foundation of Korea (NRF) Grant NRF-2021R1C1C1005076, RS-2026-25484206, the BK-21 FOUR program through NRF, and the Institute of Information \& Communications Technology Planning \& Evaluation (IITP)-Information Technology Research Center (ITRC) Grant IITP-2025-RS-2024-00437284.
AB also receives support from the NRF, under grant number NRF-2020R1I1A3068803 and JP is supported by the NRF under grant number RS-2024-00394030. YT is supported by the NSF Grant PHY-2412701. YT would like to thank the Tom and Carolyn Marquez Chair Fund for its generous support and the Aspen Center for Physics (supported by NSF grant PHY-2210452). YT would also like to thank Munich Institute for Astro-, Particle and BioPhysics (MIAPbP) which is funded by the Deutsche Forschungsgemeinschaft (DFG, German Research Foundation) under Germany's Excellence Strategy – EXC-2094 – 390783311. JHK thanks APCTP, Pohang, Korea, for its hospitality during the TRP program APCTP-2026-T02, from which this work benefited.


\appendix
\numberwithin{equation}{section}

\section{Dark Matter Density Profiles} 
\label{app:SIDM_density}
In this appendix, we provide details pertaining to the numerics to obtain the various parameters in the NFW and SIDM profiles.

\subsection{NFW Parameters}
\begin{figure}[t]
\centering
\includegraphics[width=0.48\textwidth]{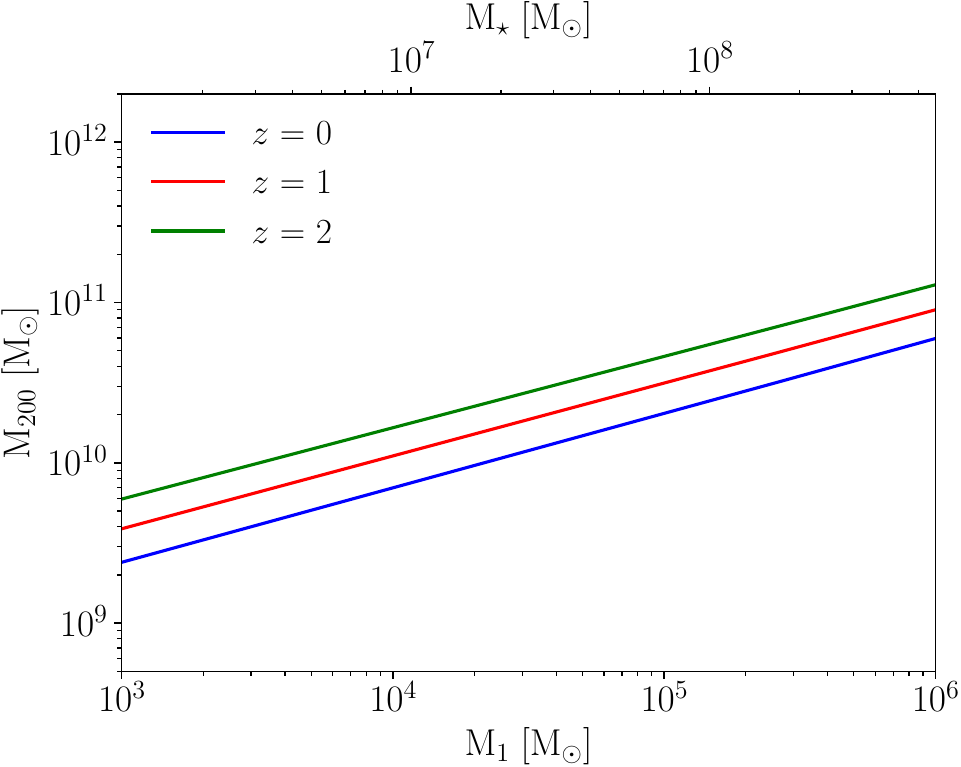}\\[2.5mm]
\includegraphics[width=0.48\textwidth]{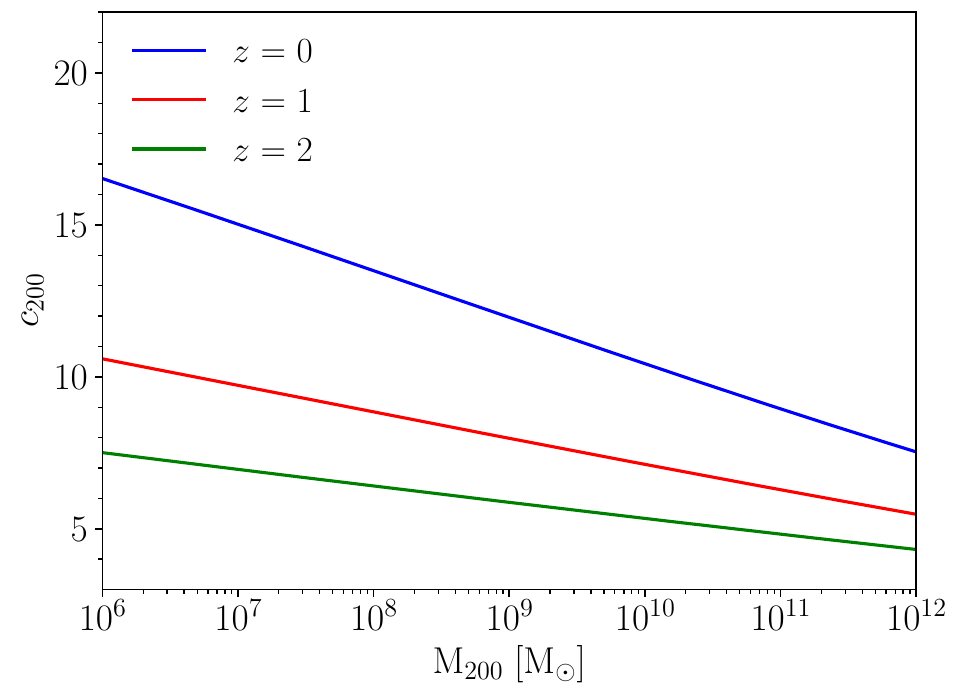}
\cprotect\caption{\textit{Top:} The halo-to-BH mass relation for different values of the redshift $z$, obtained from combining equations \eqref{eq:mstell-mBH}, \eqref{eq:mstellar} and \eqref{eq:mbulge}, where we have considered the range of masses corresponding to intermediate-mass BHs.
\textit{Bottom:} The concentration-halo mass relation \eqref{eq:cMrelation} for different redshifts.}
\label{fig:halo_relations}
\end{figure}

The starting point to obtain the parameters $\rho_{\rm sc}$ and $r_{\rm sc}$ to specify the NFW profile in \eqref{eq:rho_NFW} is to determine the viral mass $M_{200}$. For a given central BH of mass $M_1$, there exist relations involving the ``bulge'' mass $M_{\rm bulge}$ of the host galaxy, whose stellar mass is $M_*$. Concretely, the halo-to-stellar mass function is given as a function of the redshift $z$ by~\cite{Girelli:2020goz}:
\begin{equation}
    \left[M_{200}/M_*\right](z) = \frac{1}{2A(z)}\left[\left(\frac{M_{200}}{M_A(z)}\right)^{-\beta(z)} + \left(\frac{M_{200}}{M_A(z)}\right)^{\gamma(z)} \right]
    \label{eq:mstell-mBH}
\end{equation}
where the parameters $A(z), M_A(z), \beta(z)$ and $\gamma(z)$ are obtained through polynomial fitting. To relate the BH mass to the stellar mass, we first introduce the BH-bulge mass from Ref.~\cite{Kormendy:2013dxa},
\begin{equation}
    \log_{10}\left(\frac{M_1}{M_{\odot}}\right) = 8.7 + 1.1\log_{10}\left(\frac{M_{\rm bulge}}{10^{11}M_{\odot}}\right) \,.
    \label{eq:mstellar}
\end{equation}
Then, using the relation between the stellar and bulge masses~\cite{Chen:2018znx}, we obtain
\begin{equation}
    M_{\rm bulge} = k_{*,\rm bulge}\,M_*\,,
    \label{eq:mbulge}
\end{equation}
where $k_{*,\rm bulge} = 0.615$ for our cases of interest where $M_* < 10^{10}\,M_{\odot}$. In the top plot of Fig.~\ref{fig:halo_relations}, we show the final relation between the halo mass and the central BH mass.

To summarize, through the set of equations \eqref{eq:mstell-mBH}-\eqref{eq:mbulge}, one obtains the halo mass, $M_{200}$. Then, using the concentration-mass relation in \cite{Correa:2015dva},
\begin{align}
    &\left[\log_{10}c_{200}\right](z) = \nonumber \\
    &\alpha(z) +\beta(z) \log_{10}\left(\frac{M_{200}}{M_{\odot}}\right)\left[1 + \gamma(z)\log^2_{10}\left( \frac{M_{200}}{M_{\odot}}\right)\right]
    \label{eq:cMrelation}
\end{align}
where we have (different) fit functions of the redshift $\alpha,\beta$ and $\gamma$, we obtain $c_{200}$ for the corresponding halo mass. We show this relation in the bottom plot of Fig.~\ref{fig:halo_relations}. Subsequently, we obtain the NFW parameters, $r_{\rm sc}$ and $\rho_{\rm sc}$, as described in Sec.~\ref{sec:DMhalo}. 

\subsection{SIDM Parameters}

For SIDM, we need to first numerically solve the Poisson equation to determine the SIDM core profile, which we do by going along the lines of Ref. \cite{Alonso-Alvarez:2024gdz}. Assuming spherical symmetry, we have
\begin{equation}
     \left(\frac{d^2}{dr^2}+\frac{2}{r}\frac{d}{dr}\right) \ln\rho_{\rm{core}}(r) = -\frac{4\pi G}{v^2_0}\, \rho_{\rm{core}}(r) \,,
    \label{eq:SIDM_DE}
\end{equation}
subject to the following boundary conditions
\begin{equation}
    \rho_{\rm core}(r_c) = \rho_{\text{NFW}}(r_c) \equiv \rho_c\,, \quad \rho'_{\rm core}(0) = 0\,,
    \label{eqn:SIDMcore}
\end{equation}
along with the mass conservation constraint \cite{Alonso-Alvarez:2024gdz}
\begin{equation}
    \int_{r_{\rm{min}}}^{r_c}dr\,r^2\,\rho_{\rm core}(r) = \int_{r_{\rm{min}}}^{r_c}dr\,r^2\,\rho_{\text{NFW}}(r)\,.
\end{equation}
Note that $r_c$ is implicitly determined by the condition of requiring one scattering per DM particle, over the age of the core, which we re-write as (see main text)
\begin{equation}
     \frac{\sigma_0\,v_0}{m_\chi}\left(\frac{v_M}{v_0}\right)^a\,\rho_{\text{NFW}}(r_c)\,t_{\text{age}} \sim 1\,,
\end{equation}
where $a=0$ refers to contact-type interactions and for $a = 4$, we have massless-mediator type interactions. As defined in the main text, $v_M \equiv c \cdot (m_{\text{med}}/m_\chi)$. We first assume $a = 0$ to solve for the core profile. It is then convenient to switch to dimensionless variables $x \equiv r/r_c$ and $y = \ln(\rho_{\rm core}/\rho_c)$, which transforms \eqref{eqn:SIDMcore} into
\begin{equation}
    y'' + \frac{2}{x}\,y' = - Ce^{y}\,,
\end{equation}
where $C = 4\pi G \rho_c r_c^2/v_0^2$. The boundary conditions and mass conservation constraint now read as  
\begin{align}
   & \qquad \qquad  y(1) = 0\,,\quad y'(0) = 0\,,\\[1mm]
&\int_{x_{\rm{min}}}^1 dx \,x^2 e^y = \int_{x_{\rm{min}}}^1 dx \,\frac{x (1+k)^2}{(1+kx)^2}\,,
\end{align}
where $k = r_c/r_{\text{sc}}$ and $x_{\rm{min}} = r_{\rm{min}}/r_c$. For a given value of $C$, this differential equation can be solved to obtain the core profile. One may approach this as a shooting problem, i.e., by ``guessing" $y(0) = y_0$ for a fixed $C$ to match the boundary condition $y(1) = 0$. As is typical with the shooting method, there arises a degeneracy in that there may be more than one value of $y_0$ that satisfies the boundary conditions for a given value of $C$. In this case, we check that the mass conservation constraint, for a given $k < 1$, is satisfied with a tolerance of less than $0.1\%$ to accept the correct value of $y_0$. Accordingly, one obtains the dispersion velocity in the core.  

The SIDM core is then superseded by a spike profile:
\begin{equation}
 \rho^{\rm{SIDM}}_{\rm{spike}}(r) \equiv \rho_{\rm{0}} \left(\frac{r_{\rm{0}}}{r}\right)^{\gamma_{\rm{SIDM}}}\,,
 \end{equation}
where the index  $\gamma_{\rm SIDM} = 3/4$ for a contact-type interaction and $\gamma_{\rm SIDM} = 7/4$ for a massless-mediator-type interaction.
The dispersion velocity determined earlier gives the radius of the transition to the spike $r_0 = G M_1/v_0^2$, with continuity
\begin{equation}
    \rho_{\rm{core}}(r_0) = \rho_{\rm{spike}}^{\rm{SIDM}}(r_0)\,,
\end{equation}
determining the missing parameter $\rho_0$ for the SIDM spike profile.

\begin{table}[htbp]
\centering
\renewcommand{\arraystretch}{1.5}
\begin{tabular}{|c|c|c|c|}
\hline
$v_M\,[\rm{km}/s]$ 
& $\sigma_0/m_\chi\,[\rm{cm}^2/\rm{g}]$
& $r_c\,[\rm{pc}]$
& $v_0\,[\rm{km}/s]$ \\
\hline
\multirow{4}{*}{$3$}
& 2  & 14.32 & 2.49  \\
& 4  & 18.89  & 2.86  \\
& 7  & 23.63 & 3.19 \\
& 10 & 27.25  & 3.42  \\
\hline
\multirow{4}{*}{$4$}
& 2  & 19.12 & 3.02   \\
& 4  & 29.93 & 3.58   \\
& 7  & 37.44  & 4.00  \\
& 10 & 43.19 & 4.28  \\
\hline
\multirow{2}{*}{$30$}
& 2  & 19.12 & 3.02 \\
& 10 & 318.77 & 10.70  \\
\hline
\end{tabular}
\caption{SIDM core radii and dispersion velocities for the examples in Fig.~\ref{fig:density_profile} and for the analysis in Secs.~\ref{sec:inspiral} and~\ref{sec:distinguish}. For the case $v_M = 3\,{\rm km/s}$, the interaction behaves as if through the exchange of a massless mediator $(a = 4)$ for all interaction strengths, whereas for $v_M = 30\,{\rm km/s}$, the interactions are contact-type $(a=0)$ in the core and eventually transition to massless-mediator type in the spike.}
\label{tab:SIDM_params}
\end{table}

Note that we have so far assumed $v_0>v_M$. This implies contact-type interactions are valid in the core with a possible transition occurring according to the velocity dispersion ~\cite{Alonso-Alvarez:2024gdz} \footnote{We follow this analytic estimate instead of the one in Ref.~ \cite{Shapiro:2014oha}, due to requirement of continuity of the velocities between the core and spike at $r_0$.}
\begin{equation}
    v(r) = \frac{v_0}{11} \left[7 +4\left(\frac{r_0}{r}\right)^{1/2}\right]\,,
    \label{eq:vel_transition}
\end{equation}
for the contact-interaction-spike, which has $\gamma_{\rm SIDM} = 3/4$. The transition radius $r_t$ to the massless-mediator-spike is found from $v(r_t) = v_M$, and for $r<r_t$, we switch to $\gamma_{\rm SIDM} = 7/4$. Otherwise, if $v_0 < v_M$ was found, we switch to $a = 4$, and repeat the above steps to obtain the core profile and the true dispersion velocity for this case. This is then connected to the spike with exponent $\gamma_{\rm SIDM} = 7/4$, with no transition occurring, i.e., the massless mediator-type interaction is valid throughout the whole profile. In Tab.~\ref{tab:SIDM_params}, we give the values of the core radii and dispersion velocities for the various cases analyzed in this work.

Finally, we found it convenient to use the following fitting function to describe the core
\begin{equation}
    \rho^{\rm{fit}}_{\rm{core}}(r) = 
        \frac{\rho_{\rm{fit}}\, r^3_{\rm{fit}} }{(r_{\rm{fit}}+r)(r_{\rm{fit}}^2+r^2)}\,, \quad  r_0 \leq r \leq r_{\rm{fit}}\,,
\end{equation}
where the parameters $r_{\rm{fit}}$ and $\rho_{\rm{fit}}$ are determined by matching to the NFW profile at $r_{\rm{fit}}$ and to the SIDM spike profile at $r_0$.

\cprotect \section{Simulating binary-BH inspirals with \verb|KETJU|} 
\label{app:code}

\begin{figure}[t!]
\centering
\includegraphics[width=0.48\textwidth]{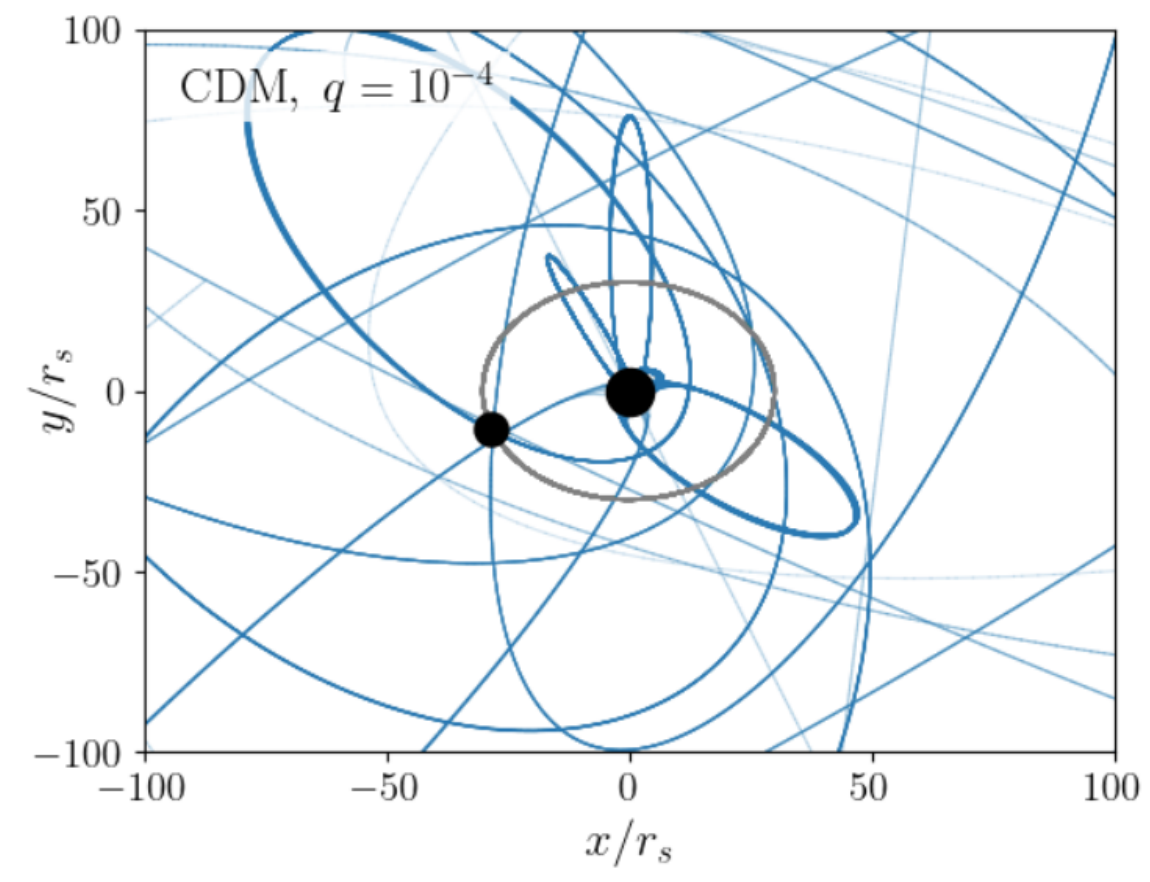} \\
\includegraphics[width=0.48\textwidth]{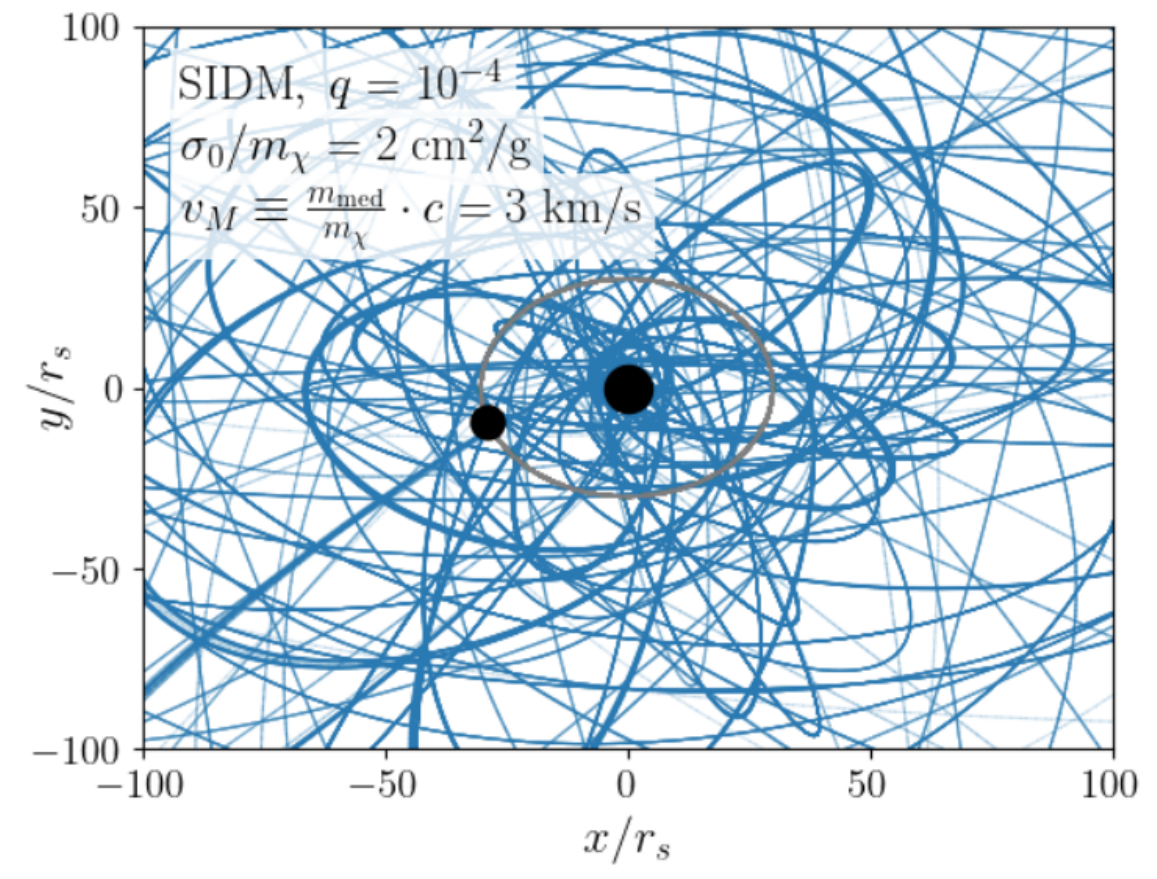}
\caption{Snapshots of our $N$-body simulation in the projected 2D plane, with the trajectories traced by the smaller BH (smaller black dot) around the central BH (larger black dot) in DM environments (blue lines) shown as gray lines, with opacity indicating increasing time. We show inspirals with $q = 10^{-4}$ in a CDM spike (top) and an SIDM spike (bottom) with $\sigma_0/m_\chi = 2\,\rm{cm^2/g}$ and $v_M = 3$~km/s. 
}
\label{fig:nbody}
\end{figure}

We provide further details here on our simulation setup. Firstly, we use the obtained density profiles to distribute the DM particles. This is done as follows: the mass of DM within a given radius is determined by integrating the density profile
\begin{equation}
    M(r) = 4\pi\int_{r_{\rm{min}}}^r du \,u^2\,\rho(u) \;.
    \label{eq:mass_profile}
\end{equation}
Using the relation $M(r) = m_{\rm{DM}}\,N(r)$, where $m_{\rm{DM}}$ is the mass of a DM clump, we then calculate the number of DM particles at radius $r$, denoted as $N(r)$. Using the density profiles, we distribute DM clumps within a non-periodic comoving box, with sizes of $\sim 1\times10^8\,r_s$  for CDM and $\sim 7 \times10^4\,r_s$ for SIDM, respectively. We choose different box
sizes to ensure similar particle numbers in the simulation, as the central density of
CDM spike is $\sim 4$ orders of magnitude larger. For CDM, we set $m_{\rm{DM}} \sim 10^{-3} M_{\odot}$, while for SIDM, we use $m_{\rm{DM}}$ in the range of $10^{-7}$--$10^{-9}\;M_{\odot}$.
A larger value of $m_{\rm{DM}}$ is chosen for CDM due to the higher density of the CDM spike profile. These choices ensure that the total number of DM clumps remains around $\sim \mathcal{O}(1000)$ in all cases, allowing for a fair comparison between CDM and SIDM. 
Simulating the modified \verb|KETJU| code with an SIDM environment for a merger time of $\tau_{\rm{merge}} \sim 4$ years requires approximately 3 days using 1 CPU core of an Intel(R) Xeon(R) Gold 5416S. In Fig.~\ref{fig:nbody}, we show snapshots of our $N$-body simulation for inspirals in CDM and SIDM environments.

For the initialization of DM velocities, we adopt spherical equilibrium initial conditions with an isotropic velocity dispersion. Firstly, we have the total gravitational potential, given by:
\begin{equation}
    \Phi(r) = \Phi_{\rm DM}(r)-\frac{GM_{\rm BH}}{r}, \qquad \Psi(r) \equiv -\Phi(r)
    \label{eq:Phi}
\end{equation}
where $\Phi_{\rm DM}$ denotes the gravitational potential of DM and the second term is the contribution of the central BH with mass $M_{\rm BH}$. Given the number density $\nu(r)\equiv \rho_{\rm DM}(r)/m_{\rm DM}$ and $\Psi(r)$, we evaluate the isotropic distribution function $f(E)$ using the Eddington inversion formula~\cite{10.1093/mnras/76.7.572}, given in its standard integral form as
\begin{equation}
    f(E) = \frac{1}{\sqrt{8}\pi^2}\int^E_0\frac{{\rm d^2}\nu}{\rm{d}\Psi^2}\frac{{\rm d}\Psi}{\sqrt{E-\Psi}},
    \label{eq:f}
\end{equation}
where $E$ denotes the binding energy. In practice, $\Psi(r)$ and $\nu(r)$ are tabulated on a logarithmically spaced radial grid, and the derivatives ${\rm d}\nu/{\rm d}\Psi$ and ${\rm d^2}\nu/{\rm d}\Psi^2$ are computed numerically on this grid. We have verified that the resulting distribution function is non-negative over the sampled energy range. Accordingly, we then sample particles from this obtained phase-space distribution as follows~\cite{Errani:2019sey}: first, clump radii are obtained through an inverse sampling method from the cumulative distribution function defined from \eqref{eq:mass_profile} as $M(r)/M(r_{\rm max})$, where $r_{\rm max}$ is determined from the condition:
\begin{equation}
    M(r_{\rm max}) = m_{\rm DM}\, N_{\rm tot}\,,
\end{equation}
where $N_{\rm tot} = 32^3$ is the total number of DM clumps considered for our simulations. Qualitatively, $r_{\rm max}$ represents the position of the last DM clump.

We then ensure that this spatial distribution matches $\rho_{\rm DM}(r)$ within the chosen radial bounds $[r_{\rm min},r_{\rm max}]$. Second, for each sampled radius $r$, we draw an energy $E \in [0,\Psi(r)]$ using rejection sampling with the conditional likelihood
\begin{equation}
    \mathcal{L}(E \mid r) \propto f(E)\,\sqrt{2\left[\Psi(r)-E\right]}\,r^{2},
    \label{eq:likelihood}
\end{equation}
which corresponds to the accessible phase-space volume at fixed radius in an isotropic system. Once $r$ and $E$ are obtained, the velocity is uniquely determined by energy conservation,
\begin{equation}
    v = \sqrt{2(\Psi(r)-E)}.
    \label{eq:v}
\end{equation}
Finally, both the position and velocity directions are assigned isotropically by drawing angles uniformly on the sphere, yielding $(x,y,z)$ and $(v_x,v_y,v_z)$ in Cartesian coordinates. This procedure produces an $N$-body realization, consistent with the target density profile and in dynamical equilibrium within the potential \eqref{eq:Phi}.

\begin{figure}[t!]
\centering
\includegraphics[width=0.48\textwidth]{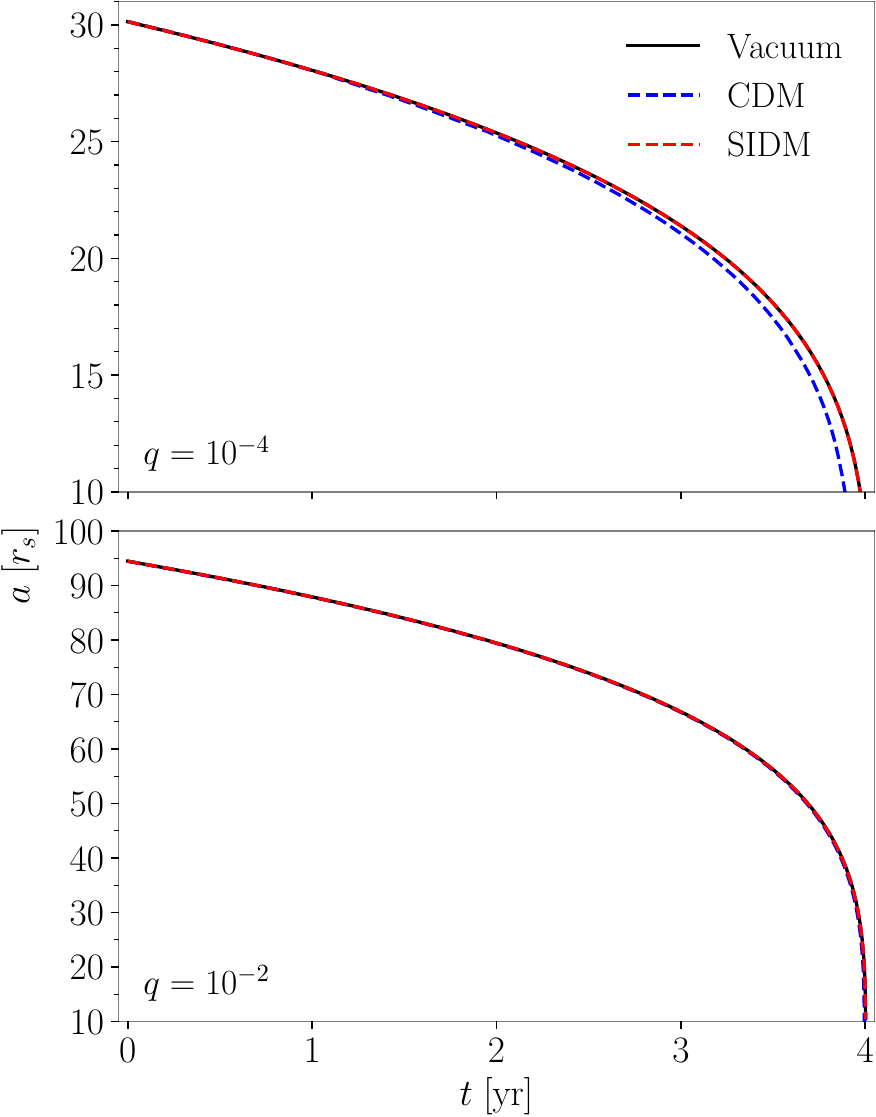} 
\caption{The evolution of the distance between the two BHs (the semi-major axis) in vacuum and DM environments for the two different mass ratios considered in this work. For SIDM, we take $\sigma_0/m = 2\,{\rm cm^2/g}$ and $v_M = 3$~km/s, which results in the highest central density amongst the various SIDM spike profiles we have studied. We have taken the binary systems to begin their merger at the same initial radius. In all cases, we find the GW emission to be the dominant source of power loss leading to binary merger. Although the SIDM and vacuum curves seem to be coincident, we show in the main text that LISA can distinguish between the two scenarios, see Fig.~\ref{fig:SIDM_delta}, based on the resultant dephasing (Fig.~\ref{fig:Ncyc_prob}). The binary separation decreases slowly in the initial stages of the inspiral, and rapidly drops at the final stages (the ``plunge'').}
\label{fig:a_vs_t}
\end{figure}

\begin{figure*}
\centering
\includegraphics[width=0.78\textwidth]{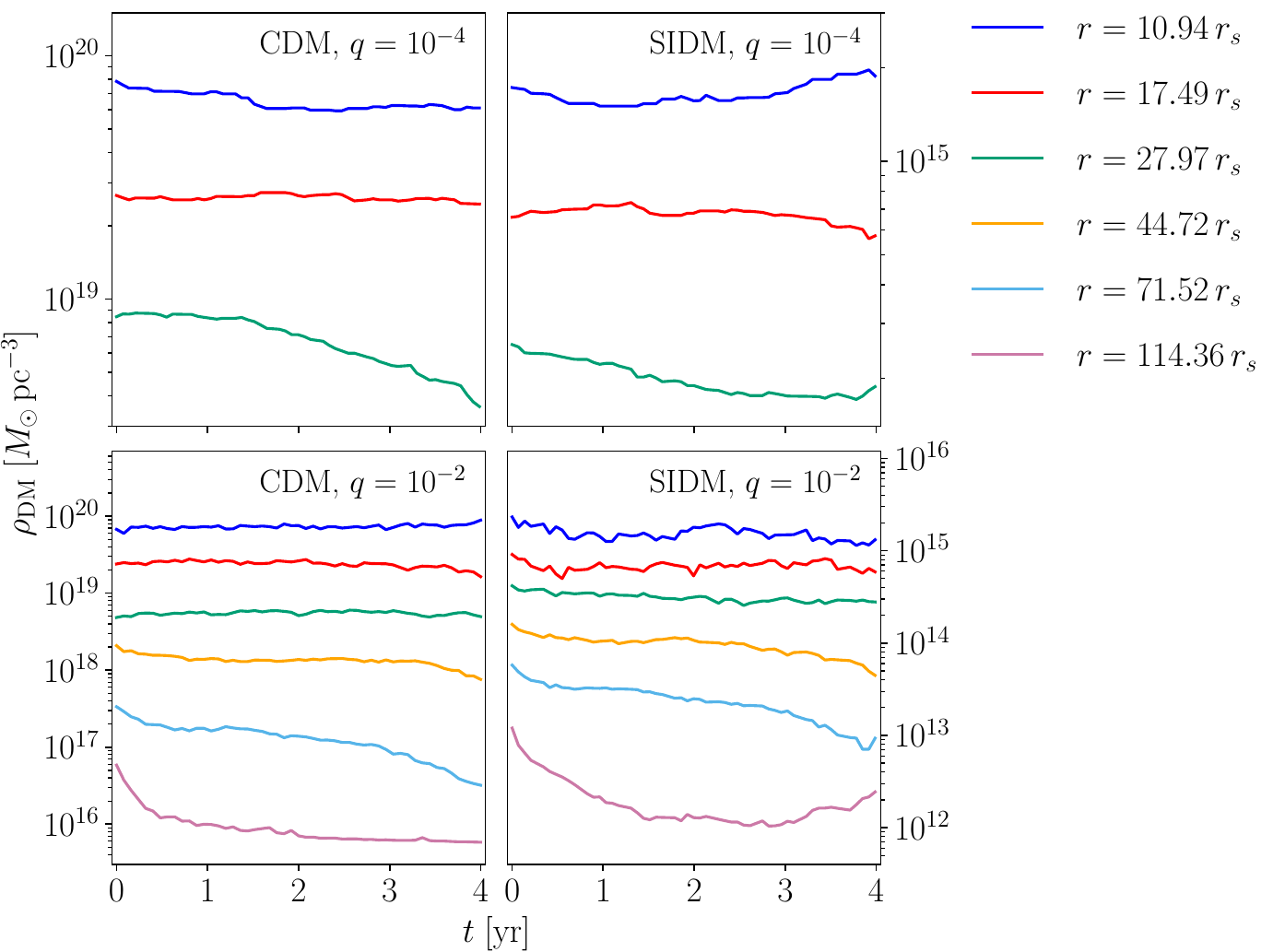} 
\caption{The evolution of densities of DM at a fixed radius, indicated by the colored lines. The drop in the density is more apparent for a larger $q$, and the depletion occurs at larger radii on account of the secondary BH spending most of the duration of the inspiral at these radii.}
\label{fig:density_fixedR}
\end{figure*}

Next, we analyze the contributions of GW emission and dynamical friction to assess their effects on orbital shrinkage within a DM environment. This comparison aims to provide a na\"{i}ve order-of-magnitude estimate using analytical formulae to the exact results obtained in Sec.~\ref{sec:inspiral}. The power emitted in GWs, at 2.5PN order, is given by \cite{Peters:1964zz},
\begin{equation}
    P_{\rm GW} = \frac{32G^4\mu^2M^3}{5c^5a^5}\frac{(1+\frac{73}{24}e^2+\frac{37}{96}e^4)}{(1-e^2)^{7/2}},
    \label{eq:Fgw}
\end{equation}
where $M \equiv M_1 (1+q)$ represents the total mass of the binary-BH system, $\mu \equiv q M_1 /(1+q)$ is the reduced mass, $a$ denotes the semi-major axis, and $e$ is the eccentricity.

The form of the dynamical friction acting on the orbiting BH through a DM medium is given by~\cite{Ostriker:1998fa}
\begin{equation}
    P_{\DF} = \frac{4\pi(G\,q M_1)^2\rho_{\text {DM}}}{v_2} I(\Mach, \Lambda)
    \label{eq:Fdf}
\end{equation}
where $v_2$ is the velocity of the lighter BH orbiting around the central BH, and the function $I$ is defined by
\begin{equation}
    I(\Mach, \Lambda)=
     \begin{cases}
      \frac{1}{2}\ln \left( \frac{1+\Mach}{1-\Mach} \right) -\Mach,   & \Mach<1, \\
     \frac{1}{2}\ln \left( 1-\Mach^{-2} \right) + \ln \Lambda ,   & \Mach>1,
    \end{cases}
    \label{eq:I}
\end{equation}
where $\Lambda\equiv ct/r_{\min}$ with $t$ representing the time for which the orbiting BH has traveled. The Mach number $\Mach\equiv v_2/c_s$ characterizes the velocity of the orbiting BH relative to the sound speed, $c_s$, in the DM medium. In the supersonic regime, specifically when $\Mach \gg 1$, the formula simplifies to the conventional steady-state result, $I= \ln (r_{\max}/r_{\min})$, also known as the Coulomb logarithm, which incorporates the maximum and minimum impact parameters associated with the orbiting BH. We adopt $\ln \Lambda = -0.5\ln q$, a commonly used approximation in the literature~\cite{2008gady.book.....B, 2010gfe..book.....M,Kavanagh:2020cfn}.
Although the exact value of the Coulomb term for SIDM varies depending on the magnitude of $c_s$ (c.f. Ref.~\cite{Fischer:2024dte} for more detailed information), this choice is sufficient for our purposes of making a naive order-of-magnitude estimate.

In Fig.~\ref{fig:a_vs_t}, we show the evolution of the semi-major axis $a$, which gives the separation distance between the two BHs. We find that GW emission is the dominant force resulting in energy loss of the binary. However, if the density of the DM spike is large enough, this results in the time to merger decreasing due to the additional energy loss, as is apparent for CDM in the top plot of Fig.~\ref{fig:a_vs_t} for $q = 10^{-4}$, resulting in a strong dephasing effect, as we have discussed in Sec.~\ref{sec:inspiral}. As mentioned before, our simulation takes into account the effects of the feedback of the secondary BH onto the DM spike, which directly impacts the orbital evolution. For larger mass ratios, this leads to a disruption of the DM halo, as shown in Fig.~\ref{fig:density_time} of the main text, leading to the evolution in DM halos resembling closer to vacuum. The disruption is significant at larger radii due to the secondary BH being present at these radii for the slower inspiral phase for a longer duration of the orbital evolution, and spending the least amount of time during the plunge phase (the end of the merger). This feature is depicted in Fig.~\ref{fig:density_fixedR} where the density near the start of the inspiral is depleted by upto an order of magnitude in $\sim 0.5$~years. Following the orbital evolution from Fig.~\ref{fig:a_vs_t}, we observe that the second BH spends $\sim 2$~years in the early stages, thereby allowing this disruption to become stronger at larger radii.

\clearpage
\onecolumngrid
\section{Fitting GW waveforms} 
\label{app:Fitting}

\begin{table}[htbp]
\centering
\renewcommand{\arraystretch}{1.5}
\begin{tabular}{|c|c|c|c|c|c|c|}
\hline
\multicolumn{3}{|c|}{\multirow{2}{*}{ENVIRONMENT}} 
& \multicolumn{2}{c|}{$|\delta M_{\rm ch}|\,[M_{\odot}]$} 
& \multicolumn{2}{c|}{$|\delta M_{\rm ch}/M_{\rm ch}|$} \\ 
\cline{4-7}
\multicolumn{3}{|c|}{} & $q=10^{-4}$ & $q=10^{-2}$ & $q=10^{-4}$ & $q=10^{-2}$ \\
\hline
\multicolumn{3}{|c|}{CDM} & $0.51$ & $0.14$ & $0.013$ & $2.23 \times 10^{-4}$ \\
\hline
\multirow{8}{*}{\begin{sideways}SIDM,~$\sigma_0/m_\chi\,[\rm{cm}^2/\rm{g}]$\end{sideways}} 
& 2  & \multirow{4}{*}{$v_M = 3\,\rm{km/s}$}  & $6.42 \times 10^{-5}$ & $3.78 \times 10^{-4}$ & $1.61 \times 10^{-6}$ &  $6.00 \times 10^{-7}$ \\
& 4  &  & $4.49 \times 10^{-5}$ & $3.35 \times 10^{-4}$ & $1.13 \times 10^{-6}$ &  $5.32 \times 10^{-7}$ \\
& 7  &  & $4.07 \times 10^{-5}$ & $3.18 \times 10^{-4}$ & $1.02 \times 10^{-6}$ & $5.05 \times 10^{-7}$ \\
& 10 &  & $4.00 \times 10^{-5}$ & $3.12 \times 10^{-4}$ & $1.00 \times 10^{-6}$ & $4.93 \times 10^{-7}$ \\
\cline{2-7}
& 2  & \multirow{4}{*}{$v_M = 4\,\rm{km/s}$}  & $4.09 \times 10^{-5}$ & $3.20 \times 10^{-4}$ & $1.03 \times 10^{-6}$ & $5.08 \times 10^{-7}$ \\
& 4  &  & $3.83 \times 10^{-5}$ & $3.08 \times 10^{-4}$ & $9.62 \times 10^{-7}$ &  $4.90 \times 10^{-7}$ \\
& 7  &  & $3.69 \times 10^{-5}$ & $3.05 \times 10^{-4}$ & $9.28 \times 10^{-7}$  &  $4.84 \times 10^{-7}$ \\
& 10 &  & $3.66 \times 10^{-5}$ & $3.02 \times 10^{-4}$ & $9.20 \times 10^{-7}$  &  $4.79 \times 10^{-7}$ \\
\hline
\end{tabular}
\caption{The mean values of the shifts in the chirp mass and the relative shift in the chirp mass when fitting the vacuum template to the environmental results.}
\label{tab:uncertainties}
\end{table}

\twocolumngrid
\bibliography{references}

\end{document}